\documentclass{PoS}

\title{Progress in Electroweak Symmetry Breaking}

\ShortTitle{Progress in EWSB}

\author{\speaker{Sally Dawson}\\
         Brookhaven National Laboratory, Upton, NY 11973 USA\\
        E-mail: \email{dawson@bnl.gov}}

\abstract{In this talk, I discuss theoretical advances in understanding the properties of the Higgs boson 
and the implications for models of electroweak symmetry breaking.  I begin by reviewing some of the recent 
progress in  Standard Model calculations for Higgs boson production and decay rates, followed by a lightning 
tour of the use of effective field theories in the search for new physics in the Higgs sector.  I end with a discussion of the complementarity of precision Higgs coupling measurements and direct searches for heavy particles for the
 discovery of Beyond the Standard Model physics in the electroweak sector.
          }

\FullConference{International Symposium on Lepton Photon Interactions at High Energies\\
		17-22 August 2015\\
		University of Ljubljana, Slovenia}

\begin{document}

\section{Introduction}

The experimental discovery of the Higgs boson marks a milestone in particle physics.  
With  the confirmation that the Higgs boson is in general consistent with Standard Model (SM)
expectations, the focus turns to precision measurements of Higgs properties and the implications for
new physics\cite{atcms}.  
The search for new physics in the electroweak sector
 depends  crucially, however, on the comparison of precision calculations with measurements of Higgs properties.
  Although no deviations from Standard Model expectations 
have been observed in the Higgs- electroweak sector as of yet, unanswered questions about the pattern of fermions masses, 
the nature of dark matter, the source of CP violation, among many other questions, lead many physicists to expect Beyond the Standard Model  (BSM) Physics at some as yet undetermined high mass scale. How this BSM physics might manifest itself at 
LHC energies is the source of considerable speculation and model building. 
BSM physics can lead to small deviations of Higgs couplings from their SM values.  In addition,
these models typically predict new Higgs-like bosons
or heavy vector resonances.  The search for these particles provides complementary information on BSM physics and is a 
major focus of Run-2 physics goals. 

\section{It Looks like the SM}

\begin{figure}[t]
\begin{centering}
\includegraphics[scale=0.37]{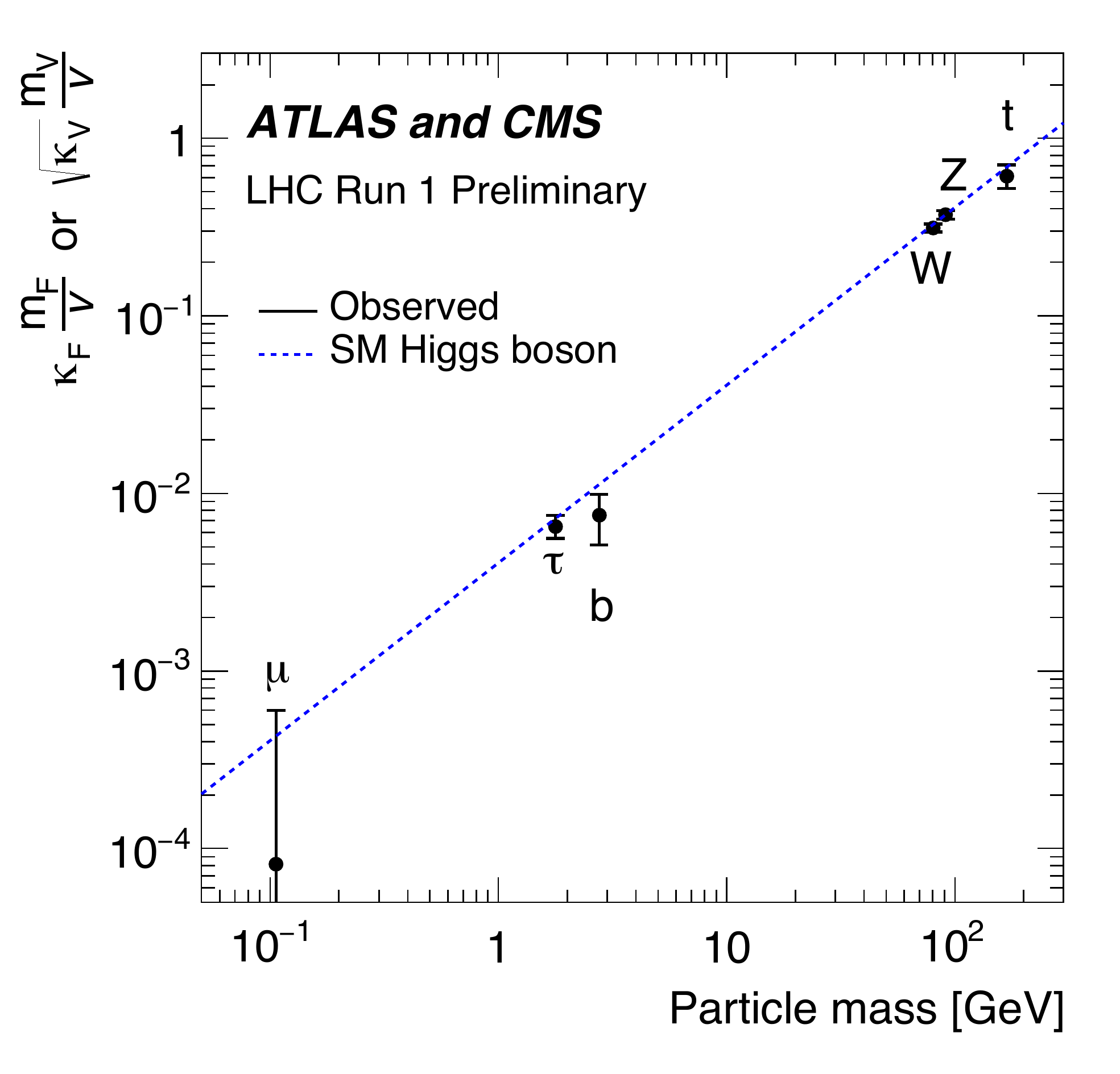}
\par\end{centering}
\caption{\em 
Relation between measured Higgs couplings and particle masses\cite{atcms}.}
  \label{fig:mass}
\end{figure}

The SM is extremely predictive, with all Higgs properties except for the mass being determined.  This
characteristic makes it a testable (and falsifiable!) model. 
Run-1 of the LHC determined that the Higgs boson is charge -$0$, parity- even, and spin-$0$, with a 
mass $m_h=125.09\pm 0.21 (stat)\pm 0.11 (syst) GeV$\cite{Aad:2015zhl}. 
Measurements of Higgs couplings are consistent with those predicted in the Standard Model at the $10-30\%$ 
level, 
and there is even
some evidence that the Higgs couplings are
proportional to fermion and gauge boson masses, as demonstrated in Fig. \ref{fig:mass}.  All of these conclusions require theoretical
calculations to the highest precision possible in the Standard Model.  Only by direct comparison of the theory with  experiment can we say that the observed particle is the SM Higgs boson.

\section{Progress in Theory Calculations}
\subsection{Higher Order QCD Corrections}

\begin{figure}[t]
\begin{centering}
\includegraphics[scale=0.6]{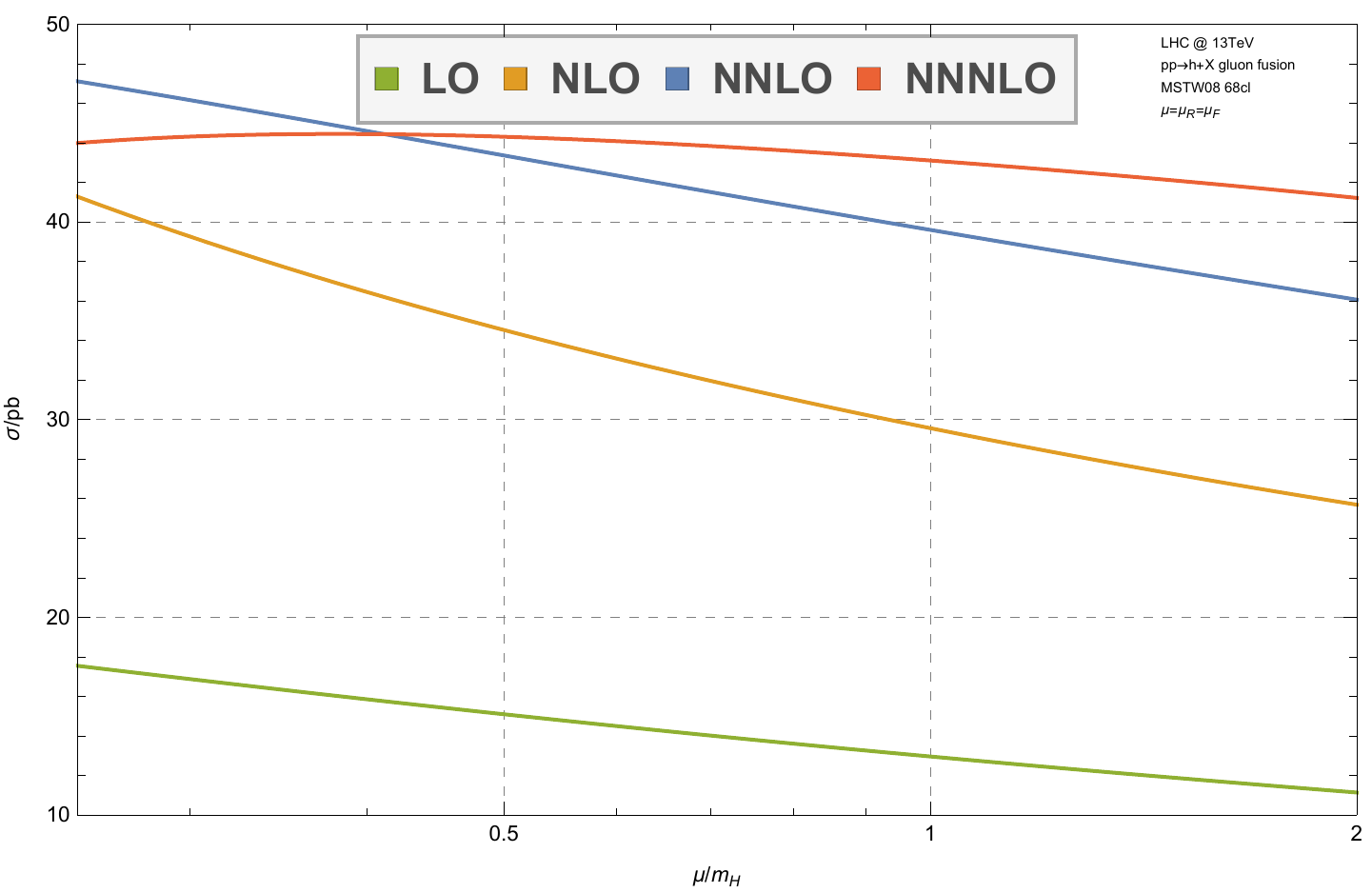}
\par\end{centering}
\caption{\em 
Scale dependence of the $gg\rightarrow h$ cross section at $\sqrt{S}=13~TeV$
at LO, NLO, NNLO, and NNNLO\cite{Anastasiou:2015ema}.
}
  \label{fig:ggh}
\end{figure}

\begin{figure}[t]
\begin{centering}
\includegraphics[clip=true, trim=0 0.9in 0 0,scale=0.4]{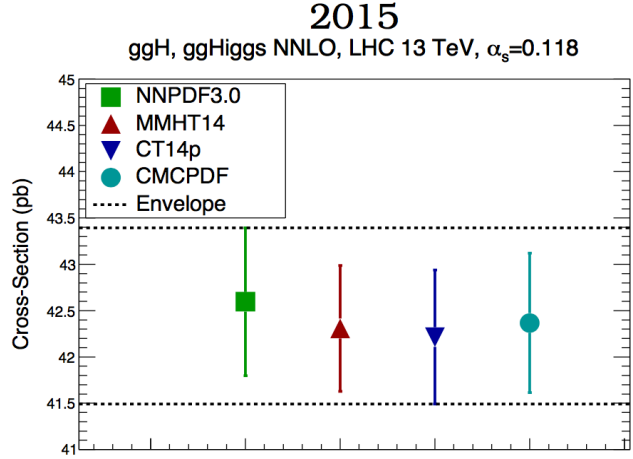}
\par\end{centering}
\caption{\em 
Dependence of NNLO Higgs cross section on the PDF choice at $\sqrt{S}=13~TeV$\cite{ff}.}
  \label{fig:gghpdf}
\end{figure}

Higgs  measurements are typically normalized relative
 to SM predictions.  This makes it of utmost importance to have reliable theory predictions.  The gold standard is becoming NNLO calculations, with many new results in the past year.
The largest 
Higgs production rate is from gluon fusion and the cross section is known at 3-loops (NNNLO) in 
the $m_t\rightarrow\infty$ limit\cite{Anastasiou:2015ema,Anastasiou:2015yha}.
The NNNLO corrections increase the rate by about $2\%$ from the NNLO prediction and reduce the scale uncertainty to $3-5\%$, as shown in 
\ref{fig:ggh}.  The most recent sets of PDFs\cite{Rojo:2015acz}
 are in remarkable agreement in predicting the NNLO Higgs rate, and the PDF uncertainty on the gluon fusion
production rate is now $\sim 2-3\%$\cite{ff}, as seen in  Fig. \ref{fig:gghpdf}.

\begin{figure}[b]
\begin{centering}
\includegraphics[scale=0.37]{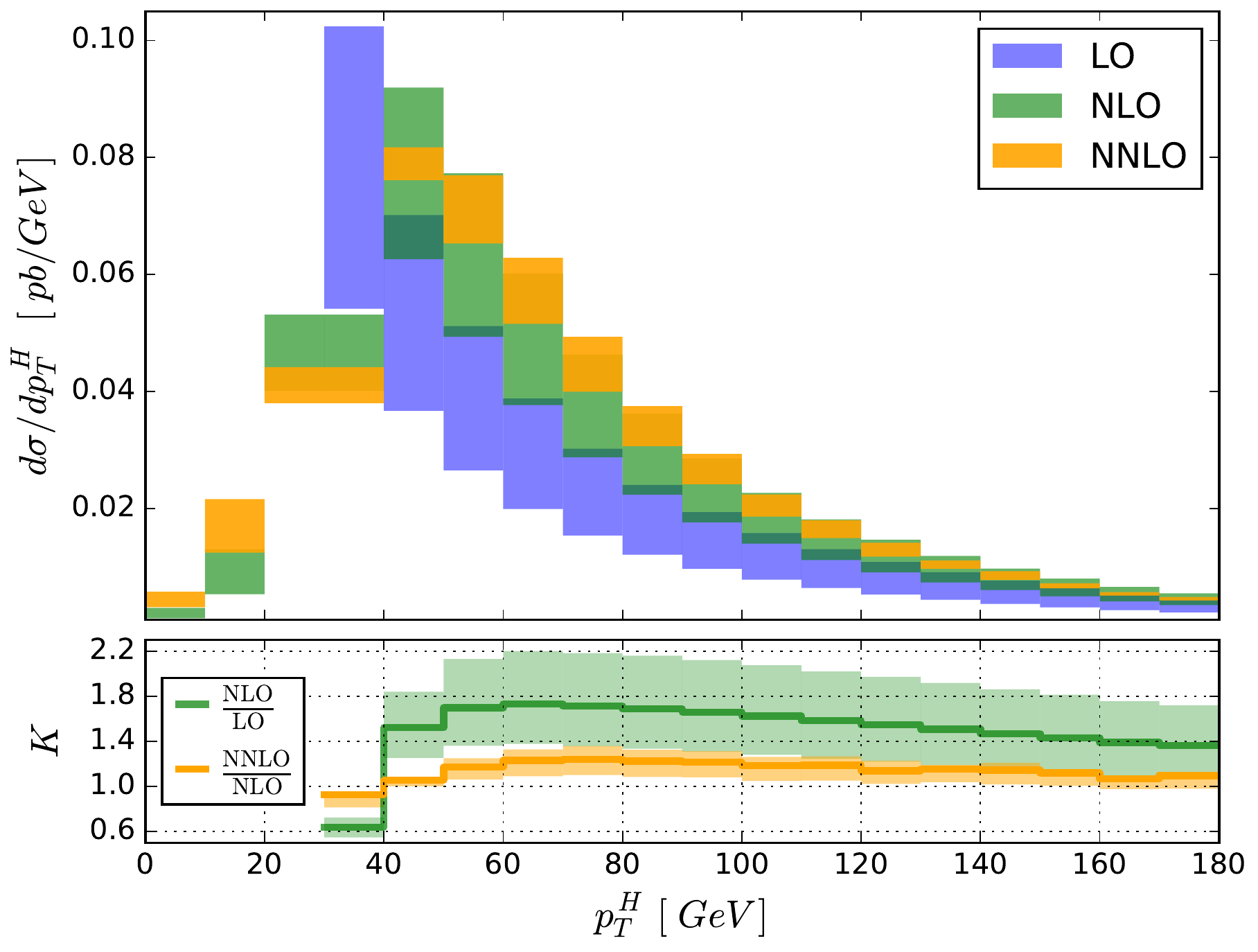}
 \hskip .2in \includegraphics[clip,height=0.45\textwidth,page=3,angle=0]{NLO-NNLO-crop.pdf}\hspace{0.8mm}%
\par\end{centering}
\caption{\em 
(a)$p_T$ dependence of the Higgs boson in $gg\rightarrow h jet$  at NNLO at $\sqrt{S}=13~TeV$\cite{Boughezal:2015aha}.
(b) $p_T$ dependence of the Higgs boson in VBF at NNLO applying VBF cuts\cite{Cacciari:2015jma}.}
  \label{fig:vbf}
\end{figure}

Searches for new physics often rely on observations in the  high $p_T$ boosted regime, 
requiring precision calculations of Higgs plus jet distributions.  Fully
differential NNLO results including all initial parton  states
 are now available\cite{Boughezal:2015aha,Boughezal:2015dra,Caola:2015wna} and are shown on the LHS 
 of Fig. \ref{fig:vbf}.  The scale dependence is reduced from the NLO prediction, and the $K$ factor has a slight
 dependence on the transverse momentum of the Higgs boson.  These calculations are also performed in the
 $m_t\rightarrow \infty$ limit. 

The rate for vector boson fusion (VBF) of a Higgs boson is significantly smaller than that from gluon fusion, but VBF can provide a clean signal
for precision measurements. VBF not only probes the Higgs couplings to gauge bosons, but eventually can be used to demonstrate that the Higgs
boson unitarizes the $WW$ scattering cross sections\cite{Lee:1977eg}.  An early calculation of VBF Higgs production at 
NNLO\cite{Bolzoni:2010xr} found 
small effects from the QCD corrections, but a new fully
differential calculation shows that when VBF  kinematic 
cuts are imposed, the NNLO corrections can be of $\sim {\cal{O}}(10\%)$\cite{Cacciari:2015jma}, as shown on the RHS of Fig. \ref{fig:vbf}.

\subsection{Measuring the Higgs Width}

\begin{figure}[b]
\begin{centering}
\includegraphics[scale=0.32,angle=-90,clip=true,trim=0 1in 0 0]{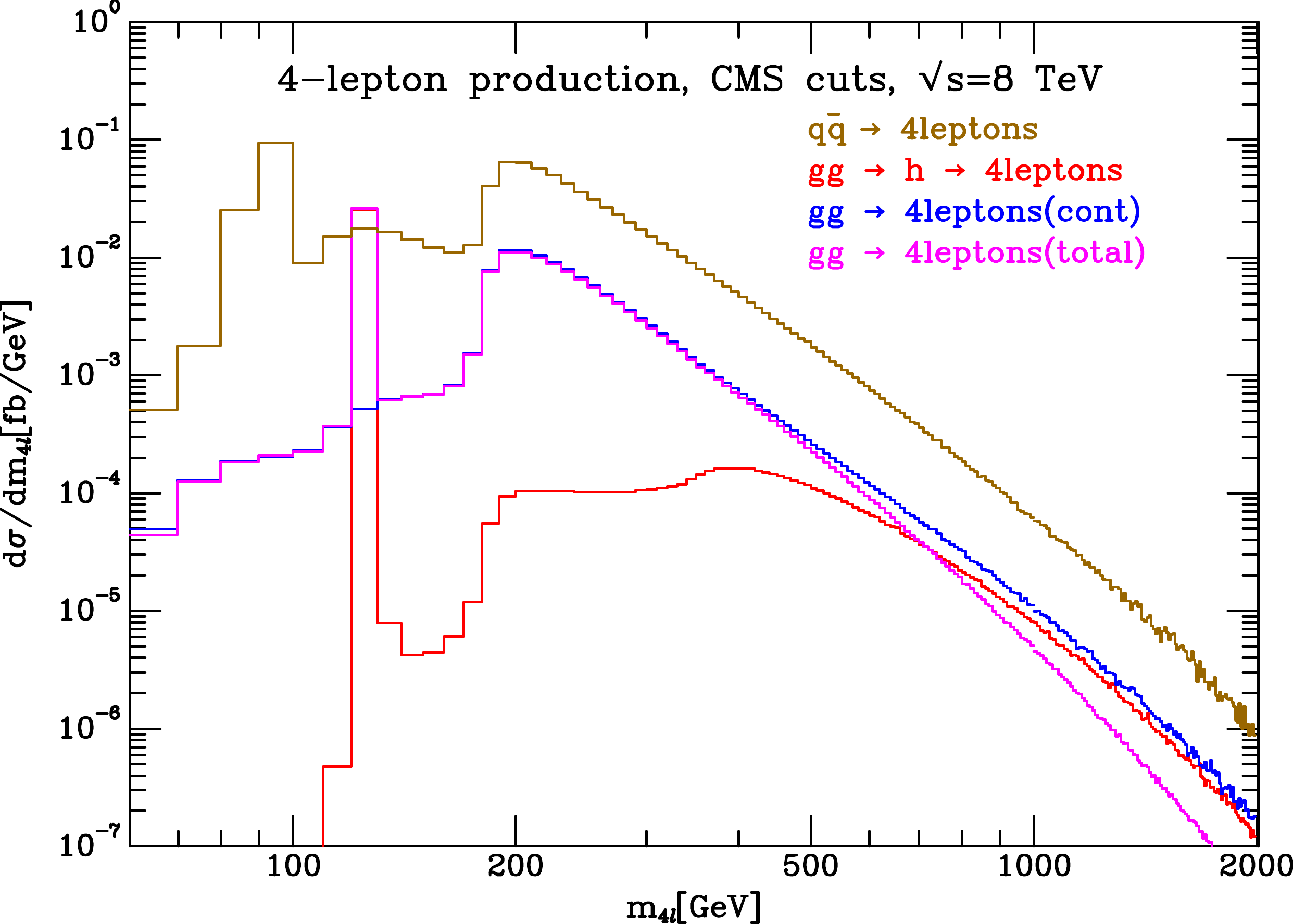}
\par\end{centering}
\caption{\em Contributions to $4$ lepton production at the LHC.  The Higgs resonance contribution is the red line.
\cite{Campbell:2013una}
}
 \label{fig:gamh}
\end{figure}
In the SM, the Higgs total width is roughly $\Gamma_h^{SM} \sim 4~MeV$ and was long assumed
 to be unmeasurable since it is so much less than
the LHC detectors' resolutions.  A few years ago, however, it was realized that
by measuring $gg\rightarrow h\rightarrow ZZ\rightarrow 4l$ on the Higgs peak,
$m_{4l}\sim m_h$,  and comparing it with measurements of $m_{4l}$ above the 
resonance peak, one obtains a  ratio  that is sensitive to the Higgs 
width\cite{Caola:2013yja,Kauer:2012hd,Campbell:2013una}.
About $15\%$ of the cross section is in the region $m_{4l}>140~GeV$, as seen in Fig. \ref{fig:gamh},  so this measurement appears feasible.  
The basic idea is that since the on-shell measurements of the Higgs cross 
section are consistent with SM expectations, a larger Higgs width would correspond to more off-shell events. Both CMS and ATLAS  have used this idea to
extract interesting limits on the Higgs width: $\Gamma_h^{ATLAS}< (4.5-7.5)\Gamma_h^{SM}$\cite{Aad:2015xua}  (with the range corresponding to different assumptions about 
unknown radiative corrections) and $\Gamma_h^{CMS}<5.5 \Gamma_h^{SM}$\cite{Khachatryan:2014iha}.  

\begin{figure}[t]
\begin{centering}
\includegraphics[scale=0.5]{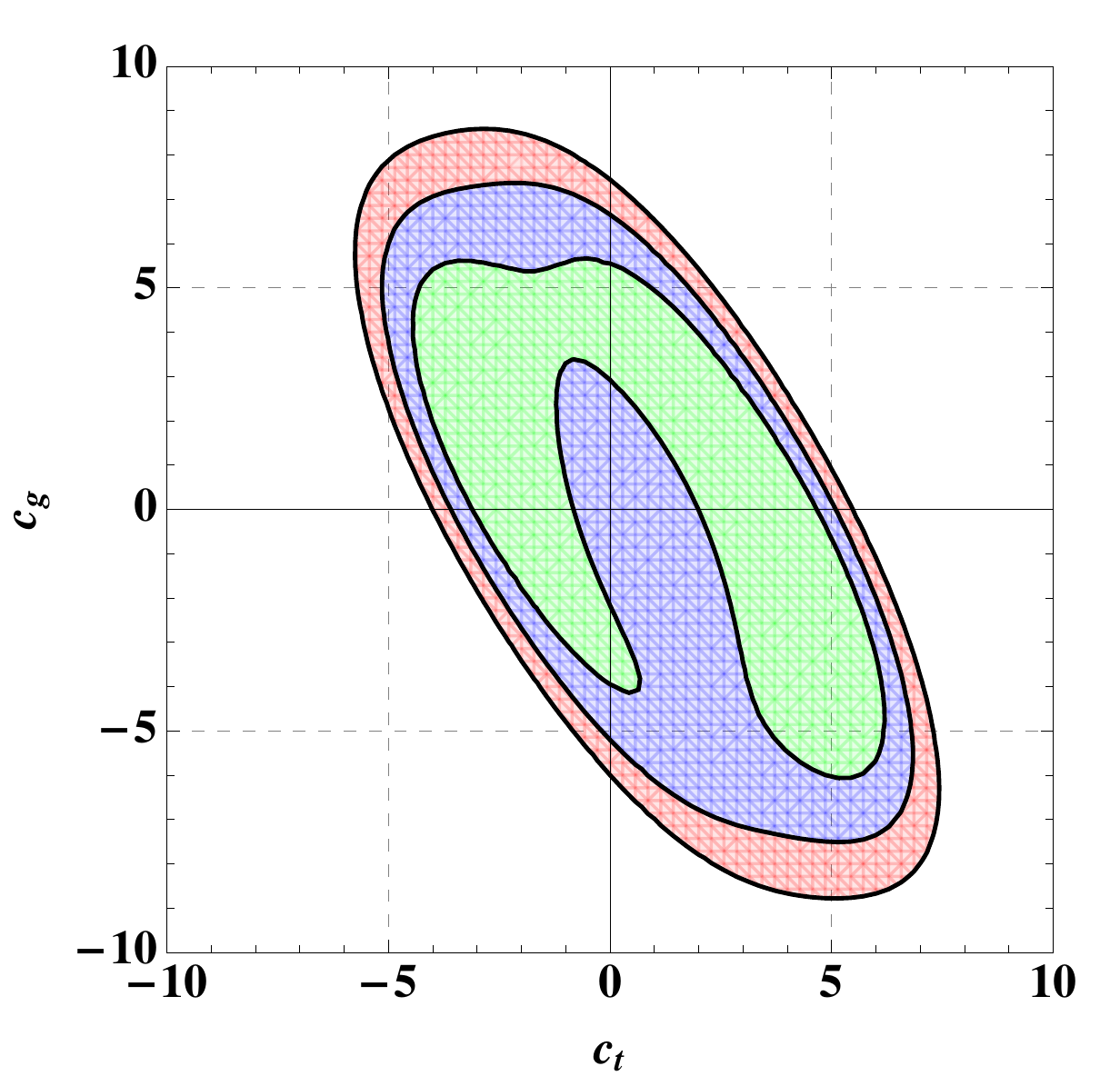}
\par\end{centering}
\caption{\em 
 $68\%$, $95\%$, and $99\%$ confidence level limits on anomalous 
 couplings from the off-shell measurement of $gg\rightarrow ZZ\rightarrow
l^+l^-l^+l^-$ using the CMS 8 TeV data\cite{Azatov:2014jga}.
}
 \label{fig:ctcg}
\end{figure}

The problem with this approach is that it assumes that Higgs couplings are the same both on the Higgs
resonance peak and above the peak.  However, this is not the case if
there is new physics (such as light colored  particles) contributing to Higgs production, or if there are non-SM contributions to the decay due to anomalous 
Higgs couplings\cite{Englert:2014ffa, Azatov:2014jga}. As an example, consider a world where all  couplings are SM- like except for the $tth$ and $ggh$ couplings,
\begin{equation}
L\sim -c_t{m_t\over v}{\overline t} t h+{\alpha_s\over 12 \pi} c_g{h\over v}G_{\mu\nu}^a G^{a.\mu\nu}\, .
\label{eq:ctcg}
\end{equation}
(In the SM,  $c_t=1$ and $c_g=0$).
The gluon fusion production rate is sensitive to $\mid c_t+c_g\mid$, while the off- shell measurements
of $gg\rightarrow ZZ\rightarrow 4l$  break this degeneracy.  Fig. 
\ref{fig:ctcg} shows the limits that  can be extracted on $c_g$ and $c_t$ from the CMS $\sqrt{S}=8 ~TeV$ measurement 
of $gg\rightarrow ZZ\rightarrow 4l$.
Imposing the restriction that $\mid c_t+c_g\mid =1$, one can extract the $68\% ~$confidence level limit $-4 < c_t < 1.5$ or $2.9 < c_t < 6.1$\cite{Azatov:2014jga}.  The limit on $c_t$ will  be improved by a direct measurement of $t {\overline t} h$ production in Run-2.

\section{Fits to Higgs Couplings}
\subsection{The $\kappa$ approximation}
One of the requirements for the observed scalar particle to be the Higgs boson of the SM is that its
couplings be those predicted by the SM.  The $\kappa$ formalism, used in analysing Run 1 Higgs data,  simply rescales
the Higgs couplings and total width from their SM values:
\begin{equation}\kappa_i={g_i\over g_{SM}}\, , \qquad \kappa_h={\Gamma_h\over \Gamma_h^{SM}}\, .\end{equation}
This approach assumes that there are no new light particles, no new tensor structures in the Higgs interactions, that the
narrow width approximation for Higgs decays is valid, and is based on rescaling total rates,
 ($\it {i.e.}$  no new dynamics is included).
The fits are done under various assumptions about new physics in loops,
 ($\it{i.e.}$  the presence of new particles contributing to the $gg\rightarrow h$ or $h\rightarrow \gamma
 \gamma$ interactions) and about the possibility of
a significant branching ratio of the Higgs to unobserved particles.  No matter how the fits are performed, the conclusion is clear:
Higgs couplings to gauge bosons and to third generation fermions must  be within $10-30\%$ of their SM predictions\cite{atcms}.

\begin{figure}[b]
\begin{centering}
\includegraphics[scale=0.35]{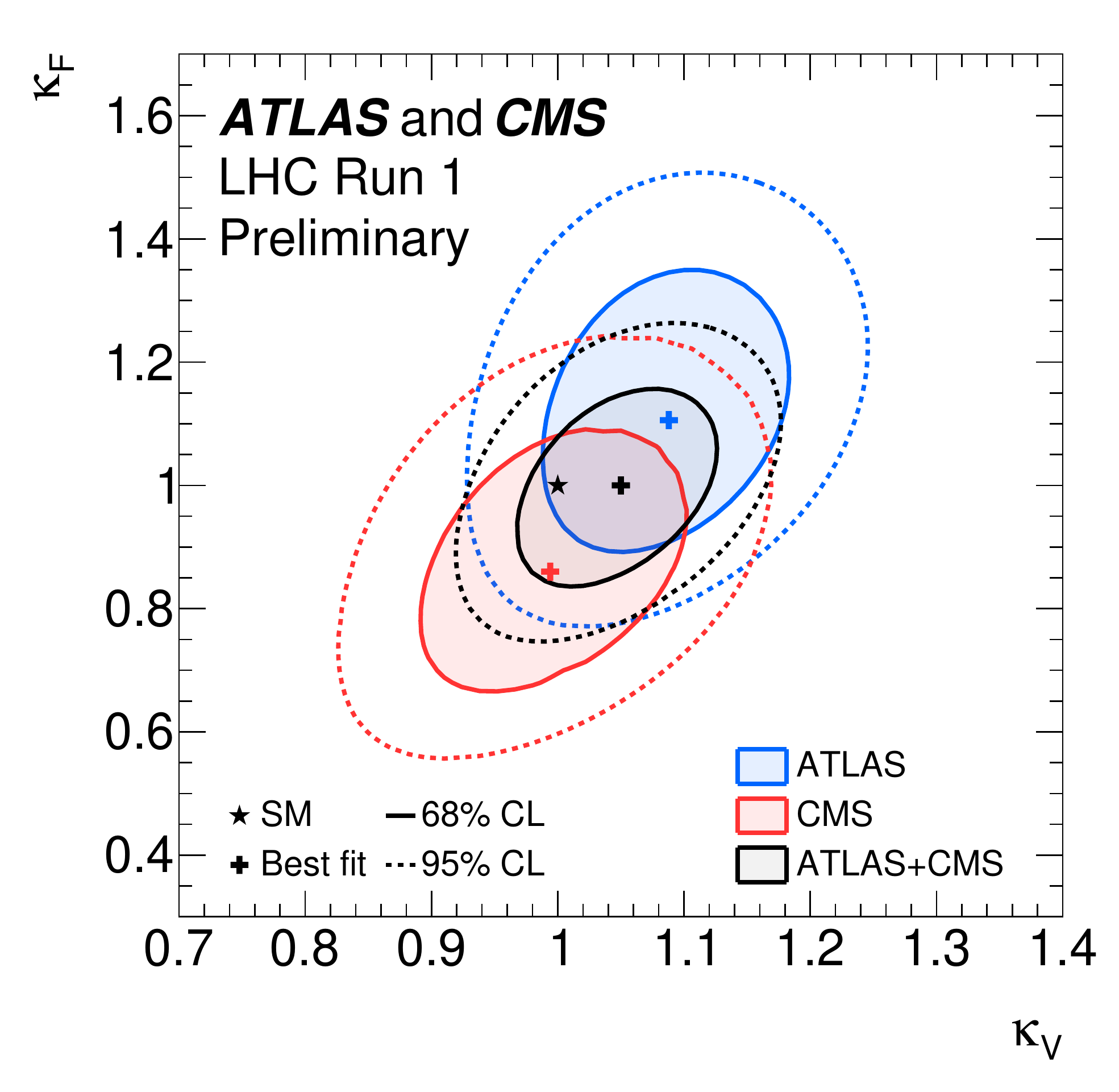}
\par\end{centering}
\caption{\em  Combined CMS and ATLAS Higgs coupling  fit when fermion and gauge boson
couplings are rescaled in a universal manner\cite{atcms}.
}
  \label{fig:higgsfit}
\end{figure}

\begin{figure}[t]
\begin{centering}
\includegraphics[scale=0.47]{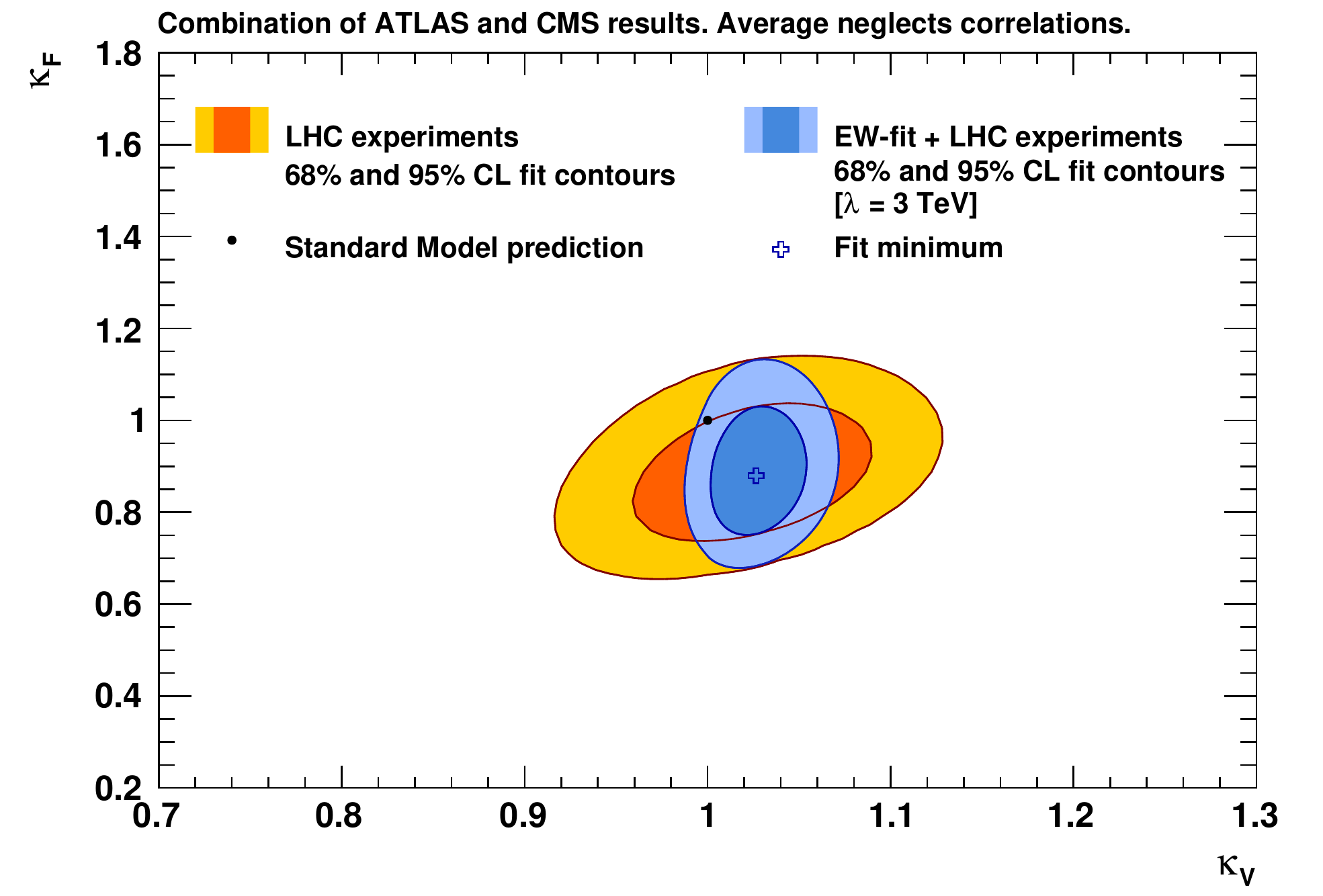}
\par\end{centering}
\caption{\em  Gfitter global electroweak fit, combined with LHC Higgs data fit when Higgs fermion and
gauge boson couplings are rescaled in an  identical fashion\cite{Baak:2014ora}.}
  \label{fig:gfit}
\end{figure}

\begin{figure}[b]
\begin{centering}
\includegraphics[scale=0.7]{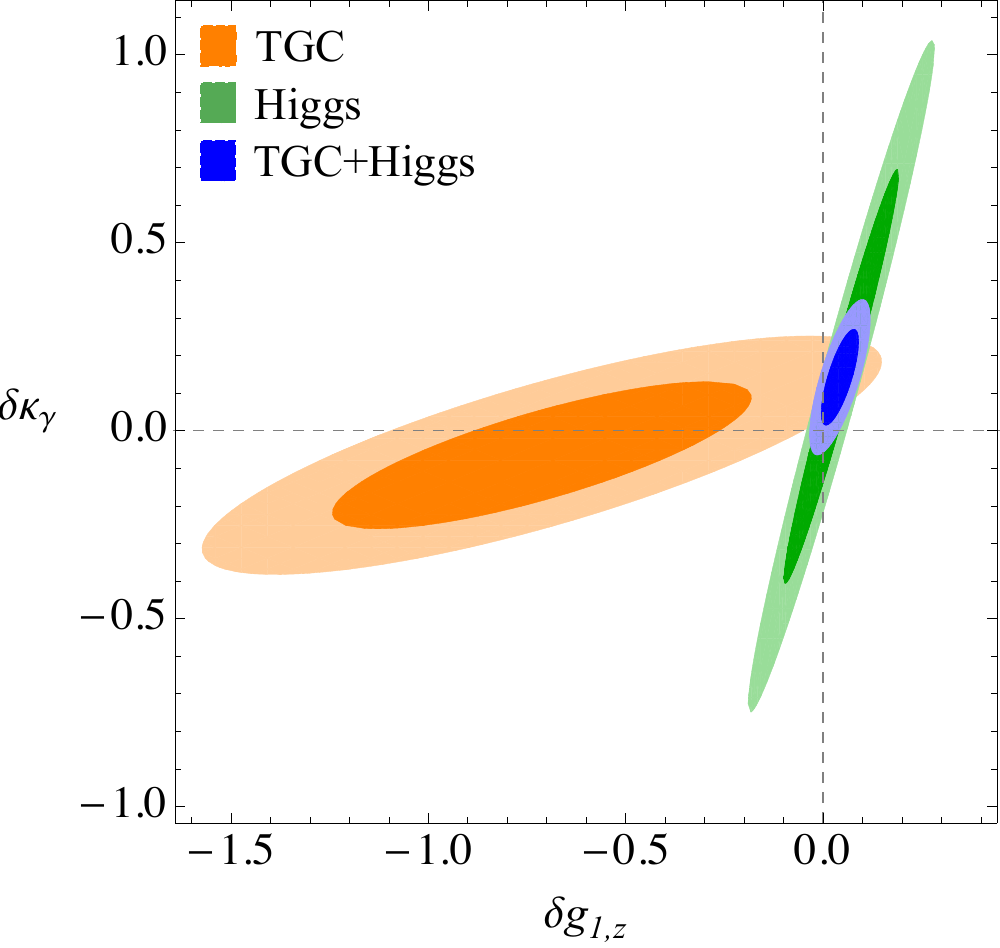}
\par\end{centering}
\caption{\em Limits from 3 gauge boson couplings (TGC) compared with limits from Higgs couplings on fits to dimension-6 EFT couplings\cite{Falkowski:2015jaa}.}
  \label{fig:3gb}
\end{figure}

A particularly simple fit can be done by assuming that all fermion and gauge boson
 couplings are rescaled in an identical fashion, 
 \begin{eqnarray}
\kappa_V&=&\kappa_W=\kappa_Z\nonumber \\
\kappa_F&=& \kappa_t=\kappa_b=\kappa_\tau=....
\end{eqnarray}
ATLAS and CMS have performed a combined fit\cite{atcms}, shown in Fig. \ref{fig:higgsfit},
where the best fit value is remarkably close to the SM value of $\kappa_V=\kappa_F=1$.  The impact of Higgs
coupling measurements on  global
electroweak fits can be seen in the Gfitter\cite{Baak:2014ora} results of Fig. \ref{fig:gfit}.    The inclusion of the electroweak data significantly strengthens the bounds obtained from Higgs coupling fits alone. 
Similar results have been found in Ref. \cite{deBlas:2014ula}. 

\subsection{Higgs Effective Field Theory }
The $\kappa$-formalism needs to be improved to analyse Run-2 data in order to incorporate
kinematic information and electroweak radiative corrections
 into the fits.   A consistent gauge invariant method to look for the effects of high mass BSM physics 
is the effective
field theory technique (EFT) in which the SM is augmented by a series of higher dimension operators, ${\cal O}_i^n$,
\begin{equation}
L=L^{SM}+\Sigma_{n>4}\Sigma_ic_i^n{O_i^n\over \Lambda^{(n-4)}}\, .
\end{equation}
The scale of new physics is generically taken to be $\Lambda$, and all operators allowed by the gauge symmetries
must be included. 
In practice, only the dimension-$6$ operators, which are suppressed by factors of $1/\Lambda^2$, are
likely to be numerically relevant.
The coefficients of the EFT are  constrained by global fits to Higgs couplings, along with precision electroweak 
measurements.   In a specific BSM model, the coefficients are calculable and are in general related to
each other. The
importance of performing a global fit is demonstrated in Fig. \ref{fig:3gb} where the constraints from Higgs data are compared with those
from the measurements of anomalous di-boson couplings\cite{Corbett:2013pja,Falkowski:2015jaa,Corbett:2015ksa}.  
The combination of Higgs data with that from 3 gauge boson vertices yields significantly stronger
constraints than using either data set alone. 
Global fits using LHC-13 data 
to the suite of dimension-6 operators will provide
significant constraints on BSM physics beyond the current limits.

\begin{figure}[t]
\begin{centering}
\includegraphics[scale=0.38]{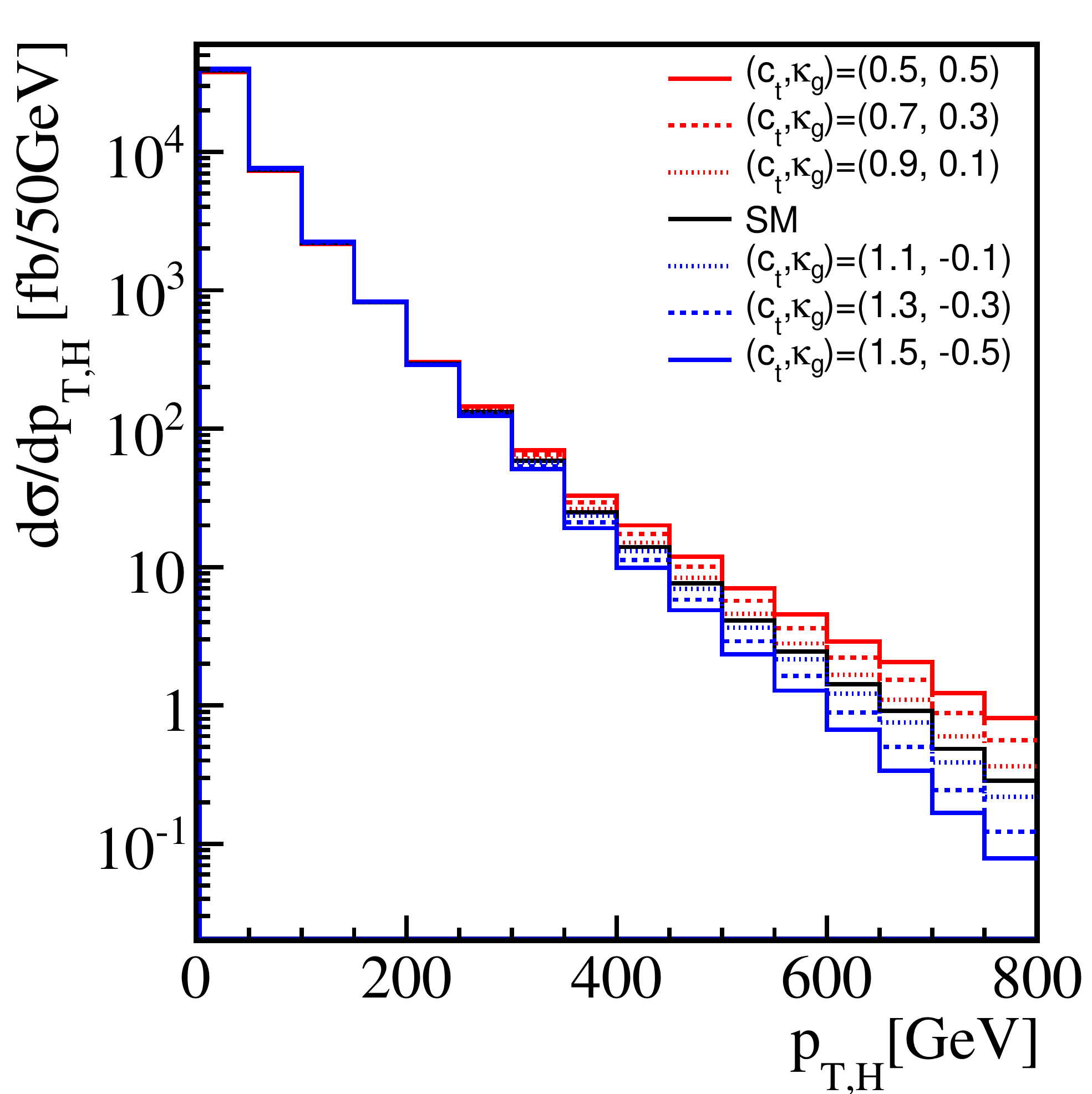}
\par\end{centering}
\caption{\em  Higgs  + jet production in the presence of anomalous couplings\cite{Schlaffer:2014osa}. The couplings are
chosen so as to leave the single Higgs production rate unchanged.}
  \label{fig:higgskin}
\end{figure}

\begin{figure}[t]
\begin{centering}
\includegraphics[scale=0.47]{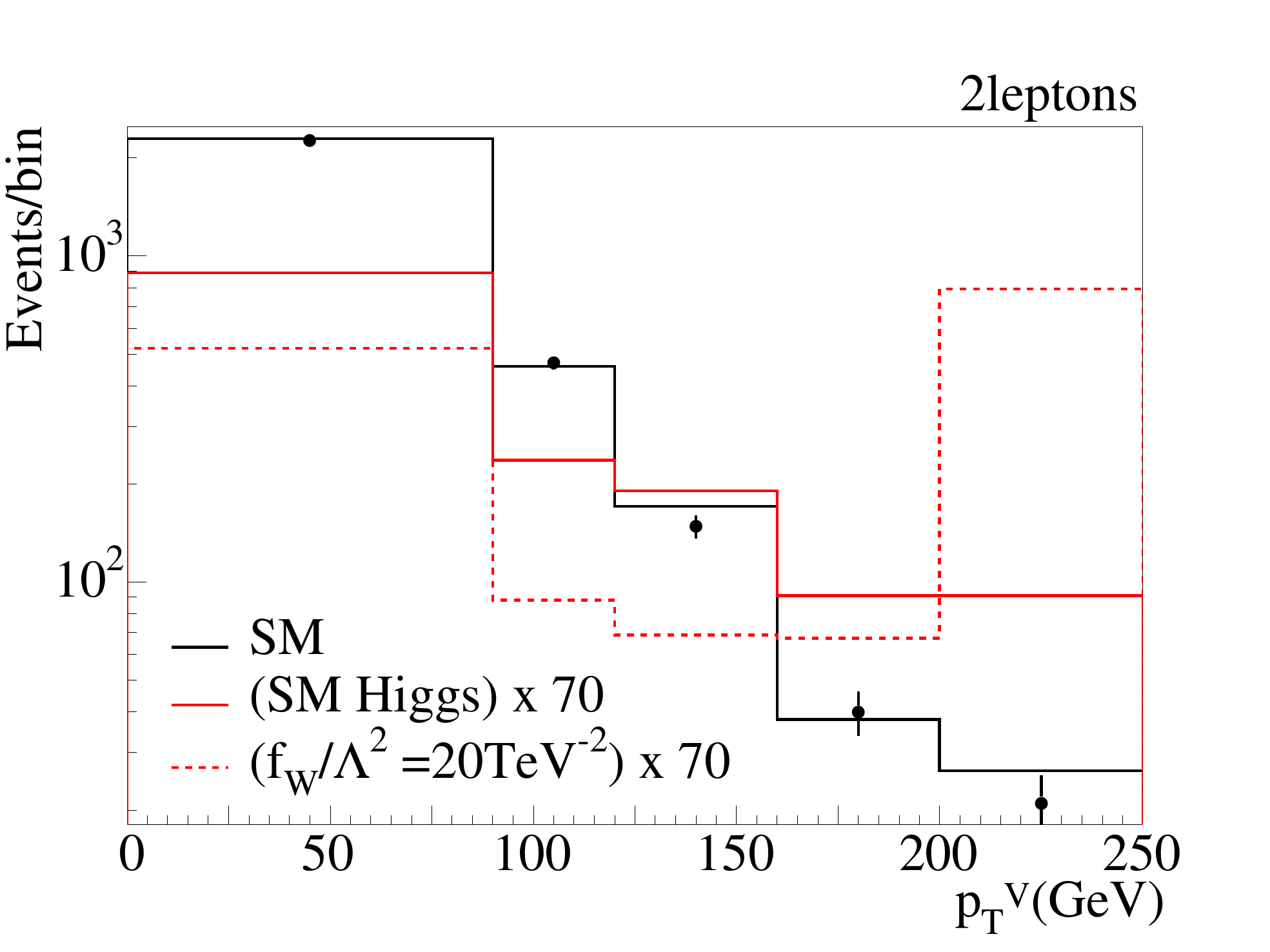}
\par\end{centering}
\caption{\em  $p_T$ distribution for $Zh, Z \rightarrow l^+l^-$ production in the presence of the anomalous coupling $f_W$ 
at $\sqrt{S}=13~TeV$\cite{Corbett:2015ksa}.}
  \label{fig:vhskin}
\end{figure}

 The dimension-$6$ operators generate contributions which typically scale like 
${p_T^2\over s}$ and so give enhanced effects in the large $p_T$ regime.   These effects are 
shown in Fig. \ref{fig:higgskin} for Higgs plus jet production, and for assumed values of $c_t$ and 
$\kappa_g\equiv c_g$ (${\it{cf}}$ Eq. \ref{eq:ctcg}).  In the tail of the distribution, the effects of the non-SM couplings become visible. 
Note that $\mid c_g+\kappa_g\mid=1$ in this plot, so the single Higgs rate is unchanged.
Fig. \ref{fig:vhskin}  demonstrates the effect of EFT couplings at high $p_T$ in 
$Zh$ production.  Note that the contribution of the EFT operator is scaled up by a factor of $70$
in order to be visible. Including information 
from kinematic distributions, therefore, can potentially lead  to
significant improvements in the fits to EFT coefficients\cite{Corbett:2015ksa}.       

\begin{figure}[b]
\begin{centering}
\includegraphics[scale=0.4]{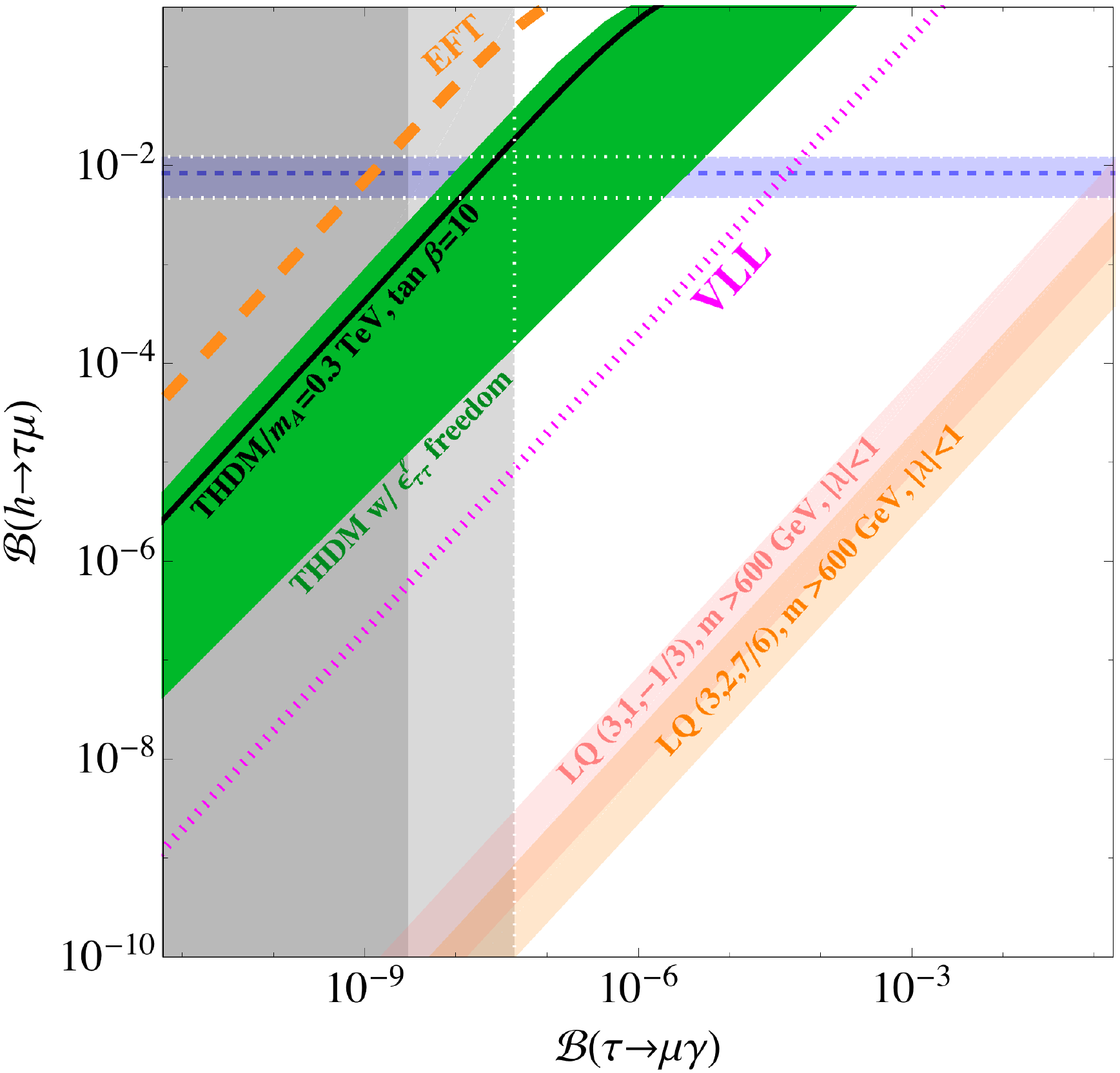}
\par\end{centering}
\caption{\em  Correlation between $BR(\tau\rightarrow \mu\gamma)$ and $BR(h\rightarrow \tau\mu)$
in various BSM models (diagonal lines)\cite{Dorsner:2015mja}.  
The horizontal band is the CMS limit on $h\rightarrow \mu
\tau$, when it is interpreted as a measurement\cite{cmslpv}.}
  \label{fig:flavor}
\end{figure}
\section{Missing Information}

Although the Higgs appears to be SM-like, we are missing
a large amount of information. 
 In particular, we know very  little about the
  Yukawa couplings of the first and second generation fermions, about CP violating  Higgs couplings, and about the flavor structure of the Higgs sector, to name just a few of our
areas of ignorance. 

\subsection{2nd Generation Yukawa Couplings}

At present there are no direct measurements of Higgs couplings to the second generation of fermions, although
there are  experimental limits on the coupling to muons which imply that $\kappa_\mu$ cannot be more than around $2$\cite{atcms}.
We also know very little about the charm quark Yukawa coupling.  If $\kappa_c$ were extremely large, the production of $c {\overline c}
\rightarrow h$ would dominate Higgs production.    A direct measurement of the charm-Higgs
coupling appears to be exceedingly difficult, leading to suggestions that the decay $h\rightarrow J/\Psi \gamma$ could potentially yield
a measurement of $\kappa_c$, since $\Gamma(h\rightarrow J/\Psi \gamma)\sim \mid 12-\kappa_c\mid^2 \times 10^{-10}~GeV$ has some
sensitivity to $\kappa_c$.
 However, the small branching ratio, $BR(h\rightarrow J/\Psi \gamma)\sim (2.8-3.0)\times 10^{-6}$ implies that
this measurement requires large luminosity\cite{Bodwin:2013gca,Koenig:2015pha}.  

\subsection{Flavor}
Experimental  limits on  flavor in the Yukawa sector come from Higgs branching ratios,
from searches for rare top decays, and low energy measurements.  In the SM, the Higgs couplings are 
diagonal  in flavor space, but it is straightforward to construct a model where this is
not the case,
\begin{equation}
L\sim \lambda^{ij}  {\overline \Psi}_L^i  \Phi b_R^j +{\tilde \lambda}^{ij}  {\overline \Psi}_L^i 
{\tilde  \Phi} t_R^j 
+{c\over \Lambda^2}(\Phi^\dagger \Phi) {\overline \Psi}_L^i\Phi b_R^j
+{c^\prime \over \Lambda^2}(\Phi^\dagger \Phi) {\overline \Psi}_L^i
{\tilde \Phi} t_R^j + 
h.c.\, ,
\end{equation}
where $i,j$ are flavor indices, $\Psi_L^i$ are the fermion doublets and  $\Phi$ is the Higgs doublet.
In the absence of a dimension-$6$ contribution, $m_b^{ij}={\lambda^{ij} v\over\sqrt{2}}$,
$m_t^{ij}={{\tilde \lambda} ^{ij} v\over\sqrt{2}}$,  
and the mass matrices and 
Yukawa matrices
are proportional to each other and simultaneously diagonalized.  
Once $c, c^\prime\ne 0$, these relationships are broken and the theory has flavor changing Higgs couplings, such as
$t\rightarrow c h$, $h\rightarrow \mu e$, etc\cite{Harnik:2012pb}.
 Flavor changing Higgs couplings in the lepton sector also contribute to low energy
observables such as $\mu\rightarrow e \gamma$ and $\tau\rightarrow \mu\gamma$.
There are complementary limits from processes such as $BR(h\rightarrow \tau \mu$) 
and $\tau\rightarrow \mu\gamma$ as shown in Fig. \ref{fig:flavor}\cite{Dorsner:2015mja}. Interpreting the CMS
limit on $h\rightarrow \mu\tau$\cite{cmslpv} as a measurement gives the horizontal band in Fig. \ref{fig:flavor}, while the
diagonal lines are the predictions of some  models with flavor violation in the Higgs sector.   In general,
there is a rich interplay between flavor violation in the Higgs sector and flavor violation in low energy observables. 
\begin{figure}[t]
\begin{centering}
\includegraphics[scale=0.6]{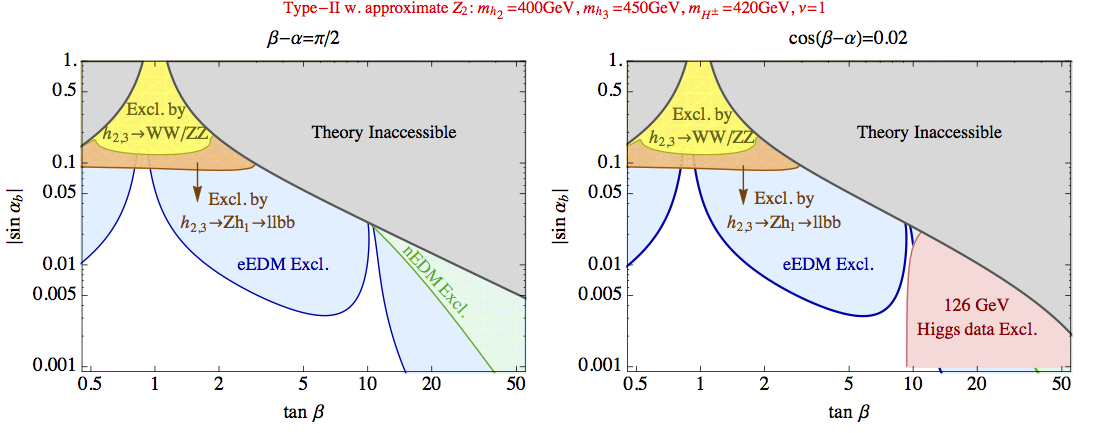}
\par\end{centering}
\caption{\em  Limits on CP violation (parameterized by non-zero $sin~\alpha_b$) in the 2HDM \cite{Chen:2015gaa}.}
  \label{fig:cp}
\end{figure}

\subsection{CP Violation in the Higgs Sector}
In the SM, all Higgs couplings are real and there is no CP violation.  The simplest way to introduce CP violation is in the
context of the 2HDM, where complex Higgs couplings are possible, leading to mixing between the $2$ neutral
scalars and the pseudoscalar boson .  These couplings change the rates for Higgs decays, 
and also change the predictions for electron and neutron EDMs\cite{Branco:2011iw,Fontes:2015xva,Chen:2015gaa}.
  A summary of current bounds is in Fig. \ref{fig:cp}, where the CP violation is parameterized by non-zero $\sin\alpha_b$.
  It is clear that the direct search for heavy Higgs bosons, the measurement of Higgs couplings, and limits from electron and neutron
  EDMs all probe complementary regions of parameter space. 

\begin{figure}[t]
\begin{centering}
\includegraphics[scale=1.]{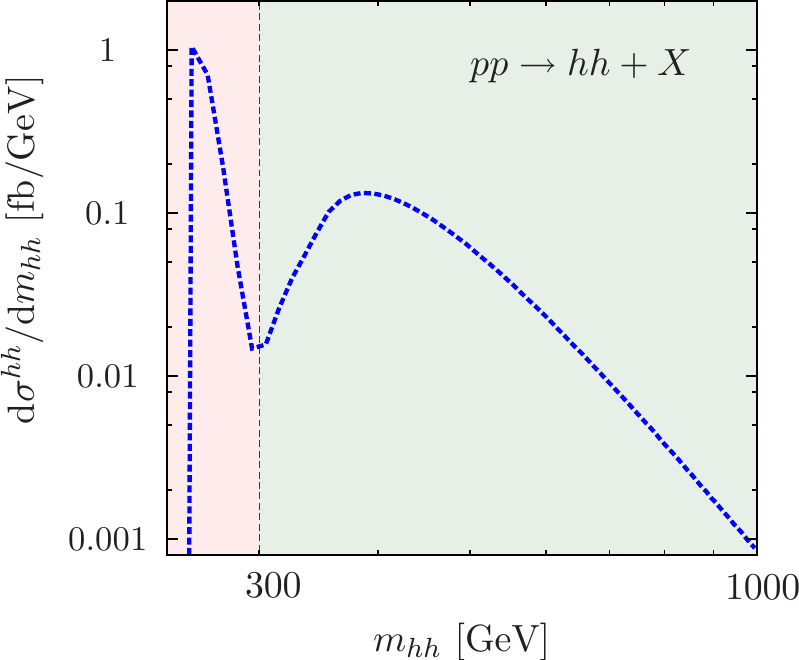}
\par\end{centering}
\caption{\em 
Higgs singlet model  prediction for $gg\rightarrow hh$ at $\sqrt{S}=14 ~TeV$ with $M_H=255~ GeV$ and 
parameters adjusted such that $\sigma/\sigma_{SM}=2.8$\cite{Dolan:2012ac}.
}
 \label{fig:gghh}
\end{figure}

\begin{figure}[b]
\begin{centering}
\includegraphics[scale=0.36]{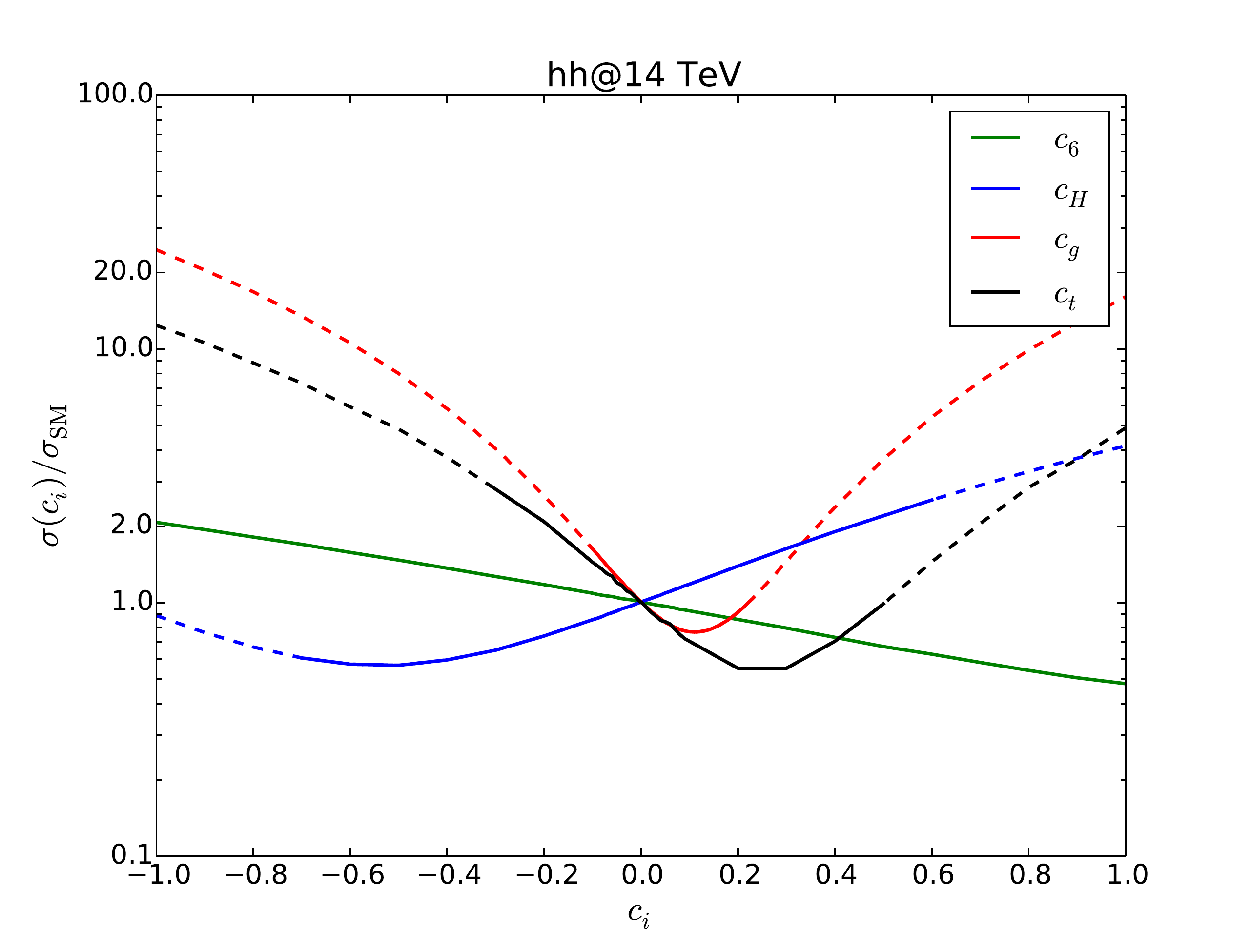}
\par\end{centering}
\caption{\em 
Sensitivity of the $gg\rightarrow hh$ cross section to anomalous couplings. 
The coefficients defined in Eq. 5.2 are normalized 
such that $c_i=0$ corresponds
to the SM.
 The dotted lines are incompatible with 
single Higgs 
measurements\cite{Goertz:2014qta}.}
 \label{fig:hhan}
\end{figure}

\subsection{Exploring the Higgs Potential}
In the minimal SM, the Higgs tri-linear coupling, $\lambda_3={m_h^2\over 2 v^2}\sim .13$, is a firm prediction of 
the model, and the measurement of $\lambda_3$ is a vital test of the structure of the SM potential.  This coupling
is probed by the process $gg\rightarrow hh$, which has an extremely small
rate in the SM: $\sigma_{NNLO}\sim 34.3~fb$ at $\sqrt{S}=13~TeV$\cite{deFlorian:2015moa}, where the NNLO rate is known in the $m_t\rightarrow
\infty$ approximation. 
The small rate makes double Higgs production quite sensitive to new physics effects, in particular in scenarios where
a resonant enhancement from a $2^{nd}$ neutral Higgs particle, $H$, is possible.  In cases where $M_H\sim 250-500~GeV$, 
enhancements of the rate by factors of up to $20$ are possible\cite{Dolan:2012ac,Robens:2015gla,Chen:2014ask,Baglio:2014nea,Hespel:2014sla}, along with distortions of the kinematic distributions, as
illustrated in Fig. \ref{fig:gghh}.  The dip in the distribution is due to interference effects between the contributions of the $2$ neutral scalars.

 Double Higgs production can also be
enhanced by factors of $2-3$ due to  anomalous
couplings \cite{Goertz:2014qta,Gillioz:2012se,Azatov:2015oxa,Chen:2014xwa}, and
the  $t {\overline t}hh$ coupling, which is typical of composite Higgs models
\cite{Panico:2015jxa}, is particularly interesting. 
Dimension-$6$ couplings
affecting double Higgs production can be parameterized as 
\begin{equation}
L\sim {c_H\over 2 v^2}
 (\partial^\mu \mid\Phi\mid^2)^2
 -{c_6\over v^2} \lambda_3 \mid\Phi\mid^6
 +{\alpha_s\over 4 \pi}{c_g\over v^2}\mid\Phi\mid^2 G_{\mu\nu}^aG^{a,\mu\nu}
-\biggl({c_t\over v^2}{\sqrt{2}m_t\over v}\mid\Phi\mid^2{\overline{\Psi_L}}{\tilde \Phi} t_R+h.c.\biggr)\, ,
\label{coefdef}
\end{equation}
and the effects of varying one coefficient at a time are shown in Fig. \ref{fig:hhan}.

\section{Naturalness and the Search for New Physics}
\begin{figure}[b]
\begin{centering}
\includegraphics[scale=0.7]{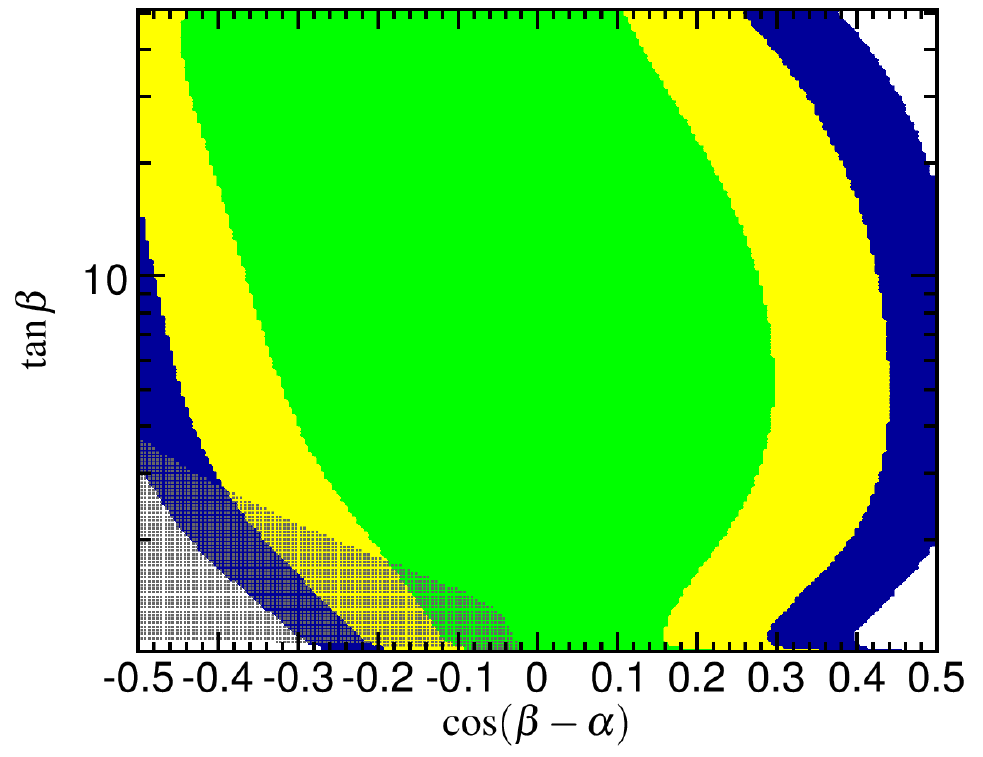}
\includegraphics[scale=0.7]{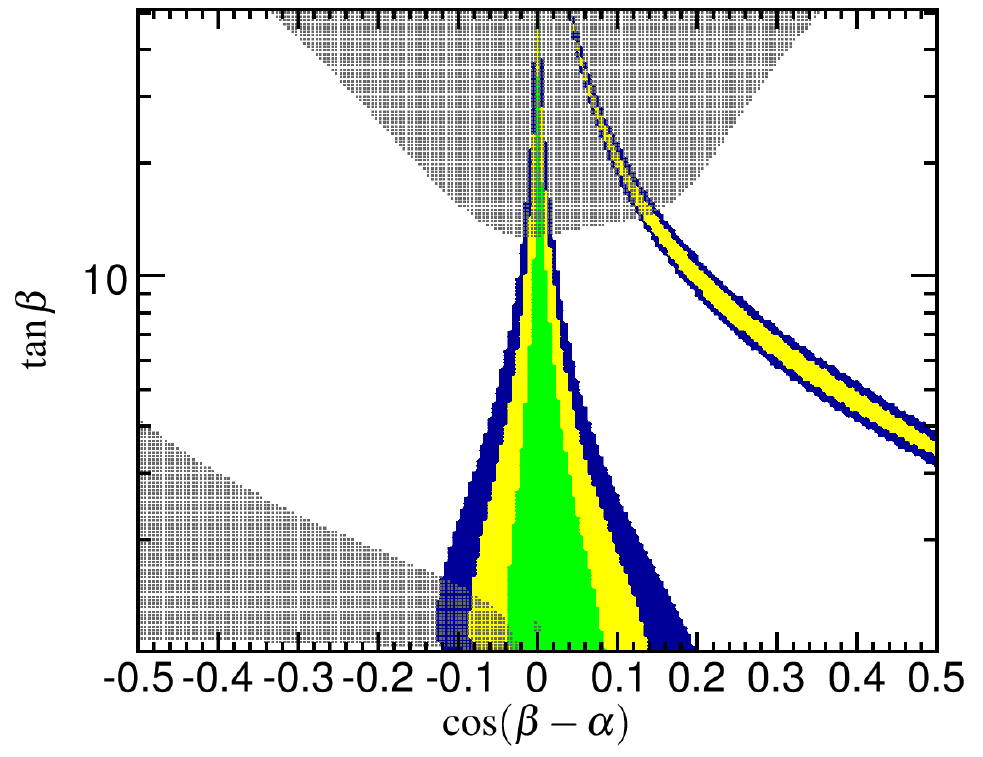}

\par\end{centering}
\caption{\em  Complementarity of limits from direct searches for the heavy Higgs boson  of a CP conserving 2HDM with $M_H=300 ~GeV$  (gray shaded) with limits from 
the precision measurements of the $125~GeV$ Higgs couplings\cite{Haber:2015pua}.
LHS:  Type-I 2HDM and RHS: Type-II 2HDM. SM couplings are obtained in the limit $\cos(\alpha-\beta)=1$.
}
 \label{fig:2hdm}
\end{figure}

Despite the impressive agreement of Higgs measurements with predictions, the theory is unsatisfactory.  In the EFT language, BSM physics generically gives contributions to the Higgs mass of ${\cal{O}}(M_h^2\sim\Lambda^2)$.
This has led to proposals for models where a symmetry prohibits large contributions to the Higgs mass
from the high scale physics--the MSSM and NMSSM are examples of this class of model. The Higgs sector of the 2HDM
illustrates many of the features of the more complicated models\cite{Haber:2010bw}.   In the 2HDM, the new physics
is probed by the search for the heavier Higgs state  and also by precision Higgs couplings.  The Higgs couplings depend on
the usual $\tan\beta$ and a mixing parameter, $\cos(\alpha-\beta)$, and there are four
possible assignments of fermion couplings
which do not lead to tree level flavor changing neutral currents.   Direct searches for the heavier Higgs boson are complementary to precision Higgs coupling measurements and both are needed to probe the parameter space.  This is illustrated in Fig. \ref{fig:2hdm}\cite{Haber:2015pua,Chen:2013rba} for two different fermion-Higgs coupling assignments. 

\begin{figure}[t]
\begin{centering}
\includegraphics[scale=0.32]{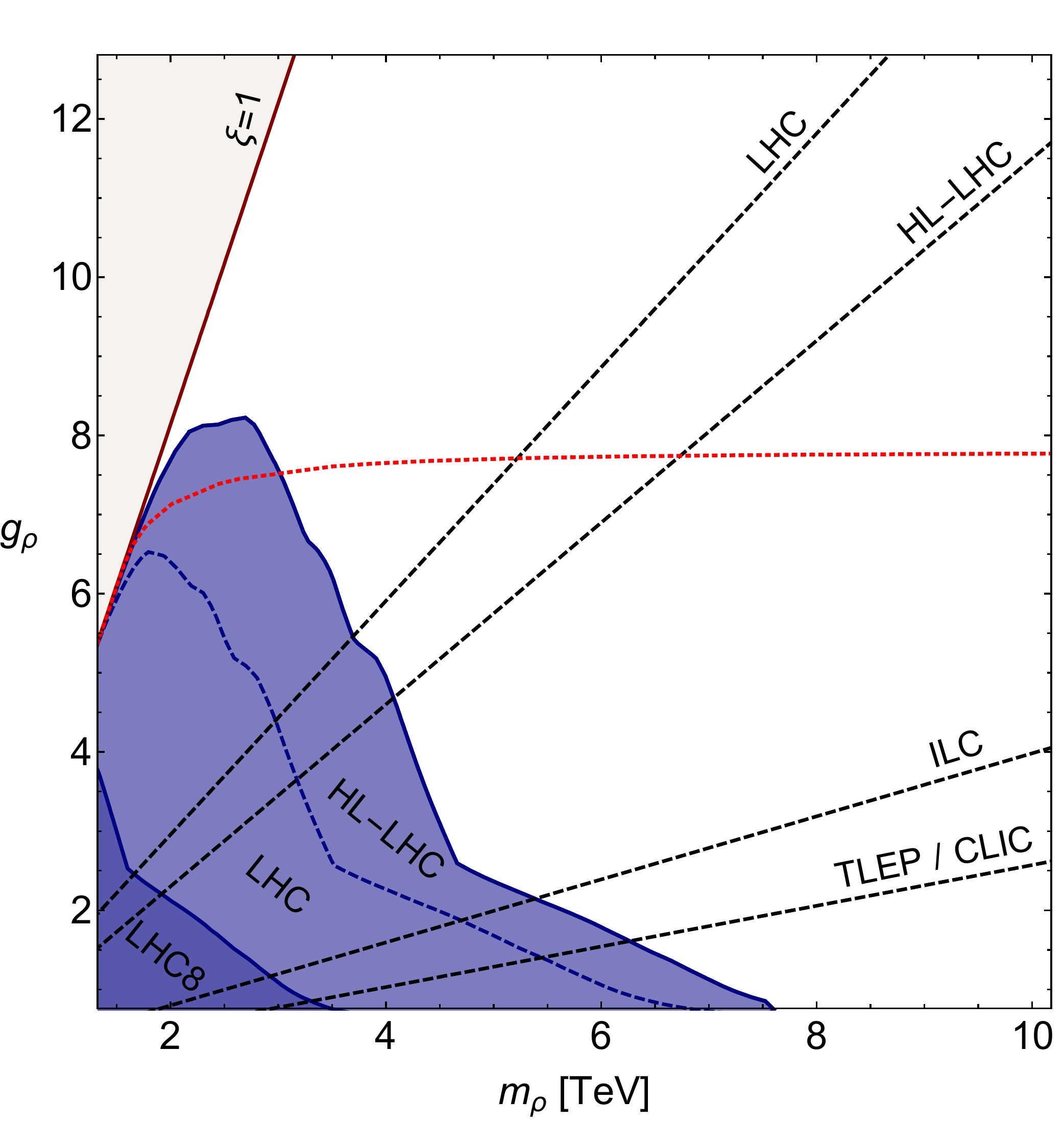}
\par\end{centering}
\caption{\em  Complementarity of direct searches for a heavy vector resonance of a composite Higgs model compared
with limits extracted from Higgs coupling measurements.  
The dotted lines are projected limits from Higgs coupling measurements
at the LHC ($\sqrt{S}=14~ TeV$ with $300 ~fb^{-1}$) and 
ILC ($\sqrt{S}=500~ GeV$ with $500~ fb^{-1}$). 
Above the red line $\Gamma_\rho/m_\rho>20\%$.\cite{Thamm:2015zwa}
}
 \label{fig:comp}
\end{figure}

An alternative possibility for solving the problem of large contributions to the Higgs mass from BSM high scale physics is
the composite Higgs scenario, where the observed Higgs particle is a pseudo Nambu Goldstone boson\cite{Panico:2015jxa}.
These models contain heavy vector resonances which are limited by direct search results. 
  Again, complimentary
limits are found from precision measurements of Higgs couplings and from  the direct searches,
as seen in Fig.\ref{fig:comp}. 

\section{Outlook}
The past year has seen immense progress in the study of electroweak symmetry breaking.  Combined ATLAS/CMS measurements
of Higgs properties, along with NNLO calculations allow for precision tests of the SM paradigm and
test BSM physics at the $TeV$ scale.  These limits are often complementary to those obtained from
low energy observables.  Run-2 will probe even
higher scales of new physics both by direct searches for new particles and by precise measurements of EFT
couplings.  It is possible that new physics lies just around the corner!

\section*{Acknowledgements}
This work is supported by the United States Department of Energy under Grant DE-AC02-98CH10886


\begin{thebibliography}{99}

\bibitem{Aad:2015zhl} 
  G.~Aad {\it et al.} [ATLAS and CMS Collaborations],
  Phys.\ Rev.\ Lett.\  {\bf 114}, 191803 (2015)
  [arXiv:1503.07589 [hep-ex]].

\bibitem{atcms} 
  The ATLAS and CMS Collaborations,
  ATLAS-CONF-2015-044.

\bibitem{Anastasiou:2015ema} 
  C.~Anastasiou, C.~Duhr, F.~Dulat, F.~Herzog and B.~Mistlberger,
  Phys.\ Rev.\ Lett.\  {\bf 114}, 212001 (2015)
  [arXiv:1503.06056 [hep-ph]].
  

\bibitem{Anastasiou:2015yha} 
  C.~Anastasiou, C.~Duhr, F.~Dulat, E.~Furlan, F.~Herzog and B.~Mistlberger,
  JHEP {\bf 1508}, 051 (2015)
  [arXiv:1505.04110 [hep-ph]].

\bibitem{Rojo:2015acz} 
  J.~Rojo {\it et al.},
  J.\ Phys.\ G {\bf 42}, 103103 (2015)
  [arXiv:1507.00556 [hep-ph]].

  \bibitem{ff}
  S.~Forte, Contribution to Higgs Hunting 2015 Conference.

\bibitem{Boughezal:2015aha} 
  R.~Boughezal, C.~Focke, W.~Giele, X.~Liu and F.~Petriello,
  Phys.\ Lett.\ B {\bf 748}, 5 (2015)
  [arXiv:1505.03893 [hep-ph]].
  
  \bibitem{Boughezal:2015dra} 
  R.~Boughezal, F.~Caola, K.~Melnikov, F.~Petriello and M.~Schulze,
  Phys.\ Rev.\ Lett.\  {\bf 115}, no. 8, 082003 (2015)
  [arXiv:1504.07922 [hep-ph]].
  
  \bibitem{Caola:2015wna} 
  F.~Caola, K.~Melnikov and M.~Schulze,
  arXiv:1508.02684 [hep-ph].
  
\bibitem{Lee:1977eg} 
  B.~W.~Lee, C.~Quigg and H.~B.~Thacker,
  Phys.\ Rev.\ D {\bf 16}, 1519 (1977).
  
\bibitem{Bolzoni:2010xr} 
  P.~Bolzoni, F.~Maltoni, S.~O.~Moch and M.~Zaro,
  Phys.\ Rev.\ Lett.\  {\bf 105}, 011801 (2010)
  [arXiv:1003.4451 [hep-ph]].
  
  \bibitem{Cacciari:2015jma} 
  M.~Cacciari, F.~A.~Dreyer, A.~Karlberg, G.~P.~Salam and G.~Zanderighi,
  Phys.\ Rev.\ Lett.\  {\bf 115}, no. 8, 082002 (2015)
  [arXiv:1506.02660 [hep-ph]].

\bibitem{Caola:2013yja} 
  F.~Caola and K.~Melnikov,
  Phys.\ Rev.\ D {\bf 88}, 054024 (2013)
  [arXiv:1307.4935 [hep-ph]].

\bibitem{Kauer:2012hd} 
  N.~Kauer and G.~Passarino,
  JHEP {\bf 1208}, 116 (2012)
  [arXiv:1206.4803 [hep-ph]].
  
  \bibitem{Campbell:2013una}
  J.~M.~Campbell, R.~K.~Ellis and C.~Williams,
  JHEP {\bf 1404} (2014) 060
  [arXiv:1311.3589 [hep-ph]].
  
  
\bibitem{Aad:2015xua} 
  G.~Aad {\it et al.} [ATLAS Collaboration],
  Eur.\ Phys.\ J.\ C {\bf 75}, no. 7, 335 (2015)
  [arXiv:1503.01060 [hep-ex]].


\bibitem{Khachatryan:2014iha} 
  V.~Khachatryan {\it et al.} [CMS Collaboration],
  Phys.\ Lett.\ B {\bf 736}, 64 (2014)
  [arXiv:1405.3455 [hep-ex]].
  
\bibitem{Englert:2014ffa} 
  C.~Englert, Y.~Soreq and M.~Spannowsky,
  JHEP {\bf 1505}, 145 (2015)
  [arXiv:1410.5440 [hep-ph]].
  
  
\bibitem{Azatov:2014jga} 
  A.~Azatov, C.~Grojean, A.~Paul and E.~Salvioni,
  Zh.\ Eksp.\ Teor.\ Fiz.\  {\bf 147}, 410 (2015)
  [J.\ Exp.\ Theor.\ Phys.\  {\bf 120}, 354 (2015)]
  [arXiv:1406.6338 [hep-ph]].



\bibitem{Baak:2014ora} 
  M.~Baak {\it et al.} [Gfitter Group Collaboration],
  Eur.\ Phys.\ J.\ C {\bf 74}, 3046 (2014)
  [arXiv:1407.3792 [hep-ph]].
  
\bibitem{deBlas:2014ula} 
  J.~de Blas, M.~Ciuchini, E.~Franco, D.~Ghosh, S.~Mishima, M.~Pierini, L.~Reina and L.~Silvestrini,
  arXiv:1410.4204 [hep-ph].
  
\bibitem{Corbett:2013pja} 
  T.~Corbett, O.~J.~P.~Éboli, J.~Gonzalez-Fraile and M.~C.~Gonzalez-Garcia,
  Phys.\ Rev.\ Lett.\  {\bf 111}, 011801 (2013)
  [arXiv:1304.1151 [hep-ph]].
  
\bibitem{Falkowski:2015jaa} 
  A.~Falkowski, M.~Gonzalez-Alonso, A.~Greljo and D.~Marzocca,
  arXiv:1508.00581 [hep-ph].


\bibitem{Corbett:2015ksa} 
  T.~Corbett, O.~J.~P.~Eboli, D.~Goncalves, J.~Gonzalez-Fraile, T.~Plehn and M.~Rauch,
  JHEP {\bf 1508}, 156 (2015)
  [arXiv:1505.05516 [hep-ph]].


\bibitem{Schlaffer:2014osa} 
  M.~Schlaffer, M.~Spannowsky, M.~Takeuchi, A.~Weiler and C.~Wymant,
  Eur.\ Phys.\ J.\ C {\bf 74}, no. 10, 3120 (2014)
  [arXiv:1405.4295 [hep-ph]].
  
    
\bibitem{Bodwin:2013gca} 
  G.~T.~Bodwin, F.~Petriello, S.~Stoynev and M.~Velasco,
  Phys.\ Rev.\ D {\bf 88}, no. 5, 053003 (2013)
  [arXiv:1306.5770 [hep-ph]].
  
  
  \bibitem{Koenig:2015pha} 
  M.~König and M.~Neubert,
  JHEP {\bf 1508}, 012 (2015)
  [arXiv:1505.03870 [hep-ph]].
  
\bibitem{Harnik:2012pb} 
  R.~Harnik, J.~Kopp and J.~Zupan,
  JHEP {\bf 1303}, 026 (2013)
  [arXiv:1209.1397 [hep-ph]].

\bibitem{cmslpv}
CMS Collaboration,
CMS-PAS-HIG-14-040, 2015.
  

  
\bibitem{Dorsner:2015mja} 
  I.~Dor¨ner, S.~Fajfer, A.~Greljo, J.~F.~Kamenik, N.~Ko¨nik and I.~Ni¨and¸ic,
  JHEP {\bf 1506}, 108 (2015)
  [arXiv:1502.07784 [hep-ph]].
  
\bibitem{Branco:2011iw} 
  G.~C.~Branco, P.~M.~Ferreira, L.~Lavoura, M.~N.~Rebelo, M.~Sher and J.~P.~Silva,
  Phys.\ Rept.\  {\bf 516}, 1 (2012)
  [arXiv:1106.0034 [hep-ph]].
  
\bibitem{Fontes:2015xva} 
  D.~Fontes, J.~C.~Romão, R.~Santos and J.~P.~Silva,
  Phys.\ Rev.\ D {\bf 92}, no. 5, 055014 (2015)
  [arXiv:1506.06755 [hep-ph]].
  
\bibitem{Chen:2015gaa} 
  C.~Y.~Chen, S.~Dawson and Y.~Zhang,
  JHEP {\bf 1506}, 056 (2015)
  [arXiv:1503.01114 [hep-ph]].

\bibitem{deFlorian:2015moa} 
  D.~de Florian and J.~Mazzitelli,
  JHEP {\bf 1509}, 053 (2015)

\bibitem{Dolan:2012ac} 
  M.~J.~Dolan, C.~Englert and M.~Spannowsky,
  Phys.\ Rev.\ D {\bf 87}, no. 5, 055002 (2013)
  [arXiv:1210.8166 [hep-ph]].
  
  \bibitem{Robens:2015gla} 
  T.~Robens and T.~Stefaniak,
  Eur.\ Phys.\ J.\ C {\bf 75}, 104 (2015)
  [arXiv:1501.02234 [hep-ph]].
  
  \bibitem{Chen:2014ask} 
  C.~Y.~Chen, S.~Dawson and I.~M.~Lewis,
  Phys.\ Rev.\ D {\bf 91}, no. 3, 035015 (2015)
  [arXiv:1410.5488 [hep-ph]].

\bibitem{Baglio:2014nea} 
  J.~Baglio, O.~Eberhardt, U.~Nierste and M.~Wiebusch,
  Phys.\ Rev.\ D {\bf 90}, no. 1, 015008 (2014)
  [arXiv:1403.1264 [hep-ph]].
  
  \bibitem{Hespel:2014sla} 
  B.~Hespel, D.~Lopez-Val and E.~Vryonidou,
  JHEP {\bf 1409}, 124 (2014)
  [arXiv:1407.0281 [hep-ph]].

 
\bibitem{Goertz:2014qta} 
  F.~Goertz, A.~Papaefstathiou, L.~L.~Yang and J.~Zurita,
  JHEP {\bf 1504}, 167 (2015)
  [arXiv:1410.3471 [hep-ph]]. 
  
  \bibitem{Gillioz:2012se} 
  M.~Gillioz, R.~Grober, C.~Grojean, M.~Muhlleitner and E.~Salvioni,
  JHEP {\bf 1210}, 004 (2012)
  [arXiv:1206.7120 [hep-ph]]. 
  
  \bibitem{Azatov:2015oxa} 
  A.~Azatov, R.~Contino, G.~Panico and M.~Son,
  Phys.\ Rev.\ D {\bf 92}, no. 3, 035001 (2015)
  [arXiv:1502.00539 [hep-ph]].
  
  \bibitem{Chen:2014xwa} 
  C.~Y.~Chen, S.~Dawson and I.~M.~Lewis,
  Phys.\ Rev.\ D {\bf 90}, no. 3, 035016 (2014)
  [arXiv:1406.3349 [hep-ph]].
  
\bibitem{Panico:2015jxa} 
  G.~Panico and A.~Wulzer,
  arXiv:1506.01961 [hep-ph].
 
 
\bibitem{Haber:2010bw} 
  H.~E.~Haber and D.~O'Neil,
  Phys.\ Rev.\ D {\bf 83}, 055017 (2011)
  [arXiv:1011.6188 [hep-ph]].
  
\bibitem{Haber:2015pua} 
  H.~E.~Haber and O.~Stål,
  Eur.\ Phys.\ J.\ C {\bf 75}, no. 10, 491 (2015)
  [arXiv:1507.04281 [hep-ph]].
  
\bibitem{Chen:2013rba} 
  C.~Y.~Chen, S.~Dawson and M.~Sher,
  Phys.\ Rev.\ D {\bf 88}, 015018 (2013)
  [Phys.\ Rev.\ D {\bf 88}, 039901 (2013)]
  [arXiv:1305.1624 [hep-ph]].
  
\bibitem{Thamm:2015zwa} 
  A.~Thamm, R.~Torre and A.~Wulzer,
  JHEP {\bf 1507}, 100 (2015)
  [arXiv:1502.01701 [hep-ph]].
\end{thebibliography}
\end{document}